\documentclass[twocolumn,superscriptaddress,floatfix,prl]{revtex4}
\usepackage[latin9]{inputenc}
\usepackage{esint}
\usepackage{graphicx}
\usepackage{xcolor}

\makeatletter
\@ifundefined{textcolor}{}
{%
 \definecolor{BLACK}{gray}{0}
 \definecolor{WHITE}{gray}{1}
 \definecolor{RED}{rgb}{1,0,0}
 \definecolor{GREEN}{rgb}{0,1,0}
 \definecolor{BLUE}{rgb}{0,0,1}
 \definecolor{CYAN}{cmyk}{1,0,0,0}
 \definecolor{MAGENTA}{cmyk}{0,1,0,0}
 \definecolor{YELLOW}{cmyk}{0,0,1,0}
}

\makeatother

\begin{document}

\title{Nematic resonance in the Raman response of iron-based superconductors}

\author{Yann Gallais }
\email{yann.gallais@univ-paris-diderot.fr}
\affiliation{Laboratoire Mat\'eriaux et Ph\'enom\`enes Quantiques, UMR 7162 CNRS, Universit\'e
Paris Diderot, Bat. Condorcet 75205 Paris Cedex 13, France}

\author{Indranil Paul}
\email{indranil.paul@univ-paris-diderot.fr}
\affiliation{Laboratoire Mat\'eriaux et Ph\'enom\`enes Quantiques, UMR 7162 CNRS, Universit\'e
Paris Diderot, Bat. Condorcet 75205 Paris Cedex 13, France}

\author{Ludivine Chauvi\`ere}
\affiliation{Laboratoire Mat\'eriaux et Ph\'enom\`enes Quantiques, UMR 7162 CNRS, Universit\'e
Paris Diderot, Bat. Condorcet 75205 Paris Cedex 13, France}

\author{J\"org Schmalian}
\email{joerg.schmalian@kit.edu}
\affiliation{Institute for Theory of Condensed Matter, Karlsruhe Institute of
Technology (KIT), 76131 Karlsruhe, Germany}
\affiliation{Institute for Solid State Physics, Karlsruhe Institute of Technology
(KIT), 76131 Karlsruhe, Germany}

\date{\today }
\begin{abstract}

In a fully-gapped superconductor the electronic Raman response has a pair-breaking peak at twice the
superconducting gap $\Delta$, if the Bogoliubov excitations are \emph{uncorrelated}.
Motivated by the iron based superconductors, we study how this peak is modified
if the superconducting phase hosts a nematic-structural quantum critical point. We show that, upon approaching
this point by tuning, e.g., doping, the growth of \emph{nematic correlations} between the quasiparticles
transforms the pair-breaking peak into a \emph{nematic resonance}. The mode energy is below $2 \Delta$,
and stays finite at the quantum critical point, where its spectral weight is sharply enhanced. The latter is consistent with recent experiments on electron-doped iron based superconductors and
provides direct evidence of nematic correlations in their superconducting phases.

\end{abstract}

\maketitle

\emph{Introduction.}--- Nematicity in correlated metals implies the propensity
of the interacting electron fluid to break rotational symmetry without
necessarily breaking translation symmetry \cite{Fradkin}. It is
at the heart of the phase competition in iron based superconductors (Fe SC) \cite{Fernandes14},
leading to a lattice softening \cite{Fernandes10,Yoshizawa12,Boehmer14},
transport and magnetic anisotropies \cite{Fisher10,Tanatar12,Fisher12, Kasahara12, Lu14}, signatures
in  optical \cite{Degiorgi,Fisherreview},
single particle \cite{Fisherreview,Davis,Yi,Pasupathy} and Raman \cite{Gallais13} spectroscopies in the normal
state. In the latter, the critical nematic fluctuations in the tetragonal phase
appear in the $B_{1g}$ Raman response as a low energy quasi-elastic peak.
\par
While nematicity in the normal state of Fe SC has been studied extensively,
it is unclear whether nematic fluctuations couple to
low energy charge carriers in the superconducting state.
In fact, even the existence and detailed nature of the nematic fluctuations
below $T_{c}$ is not established either experimentally or theoretically.
\par
In this Letter we demonstrate that, near a nematic quantum
critical point (QCP) which is well inside a fully-gapped superconducting phase~\cite{Fernandes12,Fernandes13},
the growth of nematic correlation transforms
the standard pair-breaking peak in the Raman $B_{1g}$ channel into a
\emph{new nematic resonance mode} at finite frequency.
In a clean system the resonance is a sharp structure below $2 \Delta$ of the electron pockets of Fe SC,
and its spectral weight increases strongly near the QCP. The latter, which is a key feature of the theory, and which cannot be understood within a pair-breaking peak scenario, is consistent with existing Raman scattering
measurements in Co doped Ba-122 \cite{Muschler,Chauviere10} and Na-111 \cite{Blumberg} systems.
This gives very strong evidence for this new mode, and, in turn, reveals that there is a substantial coupling between
Bogoliubov quasiparticles and nematic fluctuations.

\par
When compared to the normal state behavior,
the resonance in the superconducting phase can be understood as the consequence of the opening of the gap,
which shifts spectral weight to higher frequencies, thereby transforming the quasi-elastic peak into a resonance.
It is caused exclusively by electronic contributions, while renormalizations
due to the lattice are dynamically screened and are thus suppressed. This leads
to the interesting result that the Raman nematic resonance energy remains finite
even at a nematic QCP.

\par
The nematic resonance is analogous to the spin-resonance, observed in neutron scattering experiments
in Fe SC and cuprates superconductors \cite{Rossat,Hinkov,Bourges,Christianson08,Lumdsen09, Inosov10}.
In fact, there is a close formal
analogy of our theory to the one for the spin-resonance mode in
superconductors with sign changing gap near a magnetic instability \cite{Abanov99,Eschrig00}.
In both cases  a singular behavior of the imaginary part of a bare
response function, caused by the superconducting coherence factors,
is amplified by the vicinity to a QCP.
Important differences are that the singularity in our
case is not related to a sign change of the gap as small momentum excitations
are probed. For the same reason the Raman resonance is more sensitive
to nodes of the gap.

\par
\emph{Nematic resonance.}--- In the following we develop a theory
for the nematic resonance, and we argue that it
is intimately related to the quasi-elastic peak with diverging spectral weight experimentally observed in the
normal state $B_{1g}$ Raman response of electron doped Fe SC~\cite{Gallais13,Blumberg}.
The $B_{1g}$ quasi-elastic peak is a signature of the presence of critical fluctuations of the the nematic operator
\begin{equation}
\rho_{B_{1g}}\left(\mathbf{q}\right)=\frac{1}{N}\sum_{\mathbf{k}\sigma}\gamma_{B_{1g}}\left(\mathbf{k}\right)
\psi_{\mathbf{k}+\frac{\mathbf{q}}{2}}^{\dagger}\psi_{\mathbf{k}-\frac{\mathbf{q}}{2}},
\label{dwave-density}
\end{equation}
whose susceptibility can be accessed in Raman scattering experiments in $B_{1g}$ geometry~\cite{Devereaux-RMP}.
Here $\gamma_{B_{1g}}\left(\mathbf{k}\right)$ transforms as $k_{x}^{2}-k_{y}^{2}$ under the point group operations.
Note that, Raman experiments thus imply that the nematic instability has a
$d$-wave Pomeranchuk component. We take this as our starting point, and write a phenomenological
interaction $\mathcal{H}_{I} = -g/2 \sum_{{\bf q}} \rho_{B_{1g}}\left(\mathbf{q}\right)
\rho_{B_{1g}}\left(- \mathbf{q}\right)$,
with $g>0$, that can be tuned to study the Pomeranchuk-nematic QCP.
The microscopic origin of $g$, which can be due to
spin or bond-current \cite{Xu08,Fang08,Kang11, Paul11, Fernandes12} or orbital \cite{Lee09,Lv09,Chen10,Kontani12}
or charge \cite{Zhai09} fluctuations, is not relevant for the following discussion. Note that, fluctuations associated
with a Pomeranchuk instability are, by definition, intra-band terms. Furthermore, it is thought that
the $B_{1g}$ Raman response of electron-doped Fe SC is governed by the electron pockets \cite{Mazin10,Belen}. Consequently,
to simplify the discussion, by ($\psi_{\mathbf{k}}^{\dagger}$, $\psi_{\mathbf{k}}$) we imply the fermions describing the
electron pockets.
With this understanding, we decouple $\mathcal{H}_I$ by introducing the nematic bosonic Hubbard-Stratonovich
field $\phi\left(\mathbf{q}\right)$, and we get the action
\begin{equation}
\label{eq:action}
S= T \sum_{{\bf q}, \omega_n} \left(\chi_{{\rm nem}}^{0}\left({\bf q}, i \omega_n \right)^{-1}
-g\right)\left| \phi_{{\bf q}, \omega_n } \right|^2 +\cdots.
\end{equation}

The bare nematic susceptibility $\chi_{{\rm nem}}^{0}\left({\bf q}, i \omega_n \right)$ is the Fourier transform of
$\langle T_{\tau} \rho_{B_{1g}}\left(\mathbf{q},\tau \right)
\rho_{B_{1g}}\left(\mathbf{q},0\right) \rangle_0$, where $\langle \cdots \rangle_0$
implies average in the BCS ground state with $g=0$.

In Raman spectroscopy of Fe SC the momentum transfer by the photons is typically small compared to the frequency transfer
($\omega \gg v_F q$, where $v_F$ is typical Fermi velocity). Consequently, the $B_{1g}$ Raman response function probes the imaginary part of the
full nematic susceptibility $\chi_{{\rm nem}}\left({\bf q}, i \omega_n \right)$
in the dynamical limit, and, in random phase approximation, is given by
\begin{equation}
R_{B_{1g}}\left(\omega\right) \equiv {\rm Im} \chi_{{\rm nem}}\left(\omega\right)
={\rm Im}\left(\frac{\chi_{{\rm nem}}^{0}\left(\omega\right)}
{1-g\chi_{{\rm nem}}^{0}\left(\omega\right)}\right),
\label{eq:Raman}
\end{equation}
where we suppressed the momentum $\mathbf{q}=\mathbf{0}$.

\begin{figure}
\centering
\includegraphics[clip,width=0.99\linewidth]{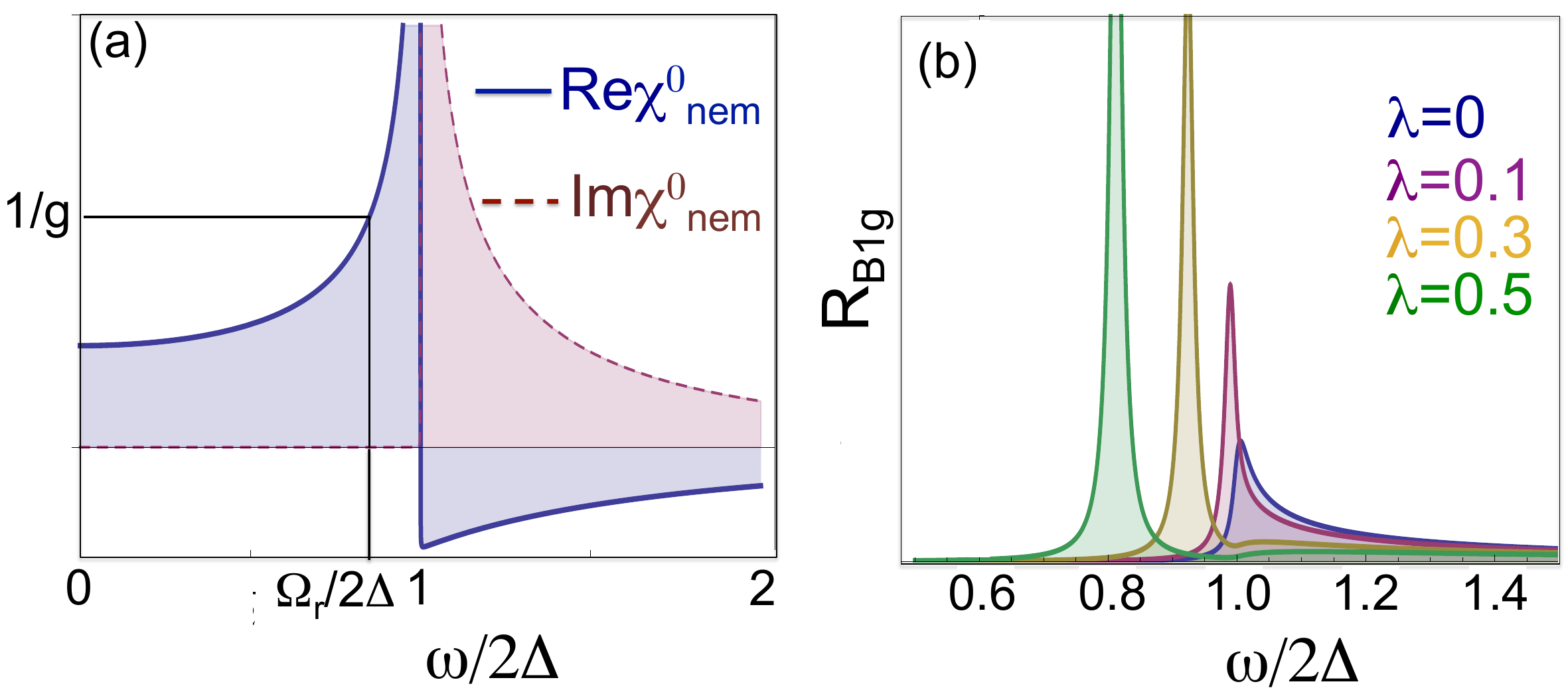}
\caption{(a) Real and imaginary parts of the bare BCS Raman response $\chi^0_{nem}$ for an isotropic gap.
The interacting response (Eq. \ref{eq:Raman} in the main text) develops a resonance at $\Omega_r <  2\Delta$ when
the real part reaches the threshold $1/g$ (see Eq. \ref{eq:Raman} in text). (b) Development of the nematic resonance
in the superconducting $B_{1g}$ Raman response $R_{B_{1g}}$ for different values of the dimensionless coupling
constant $\lambda$ (see text). Note that for $\lambda =$ 0.3, 0.5, weak pair breaking continuum are visible.}
\label{fig1}
\end{figure}

For simplicity we evaluate the above at $T=0$ and by
ignoring lifetime broadening of the Bogoliubov quasiparticles. Later we comment about finite-$T$ and
finite lifetime effects. Furthermore, we assume a constant superconducting gap $\Delta$ on the electron pockets
(for other scenarii see \cite{boyd09,SI}).
It is then straightforward to determine the imaginary part of the bare $B_{1g}$ Raman response \cite{boyd09}
\begin{equation}
{\rm Im}\chi_{{\rm nem}}^{0}\left(\omega\right)=\frac{\pi \rho}{2}\frac{\left(2\Delta\right)^{2}\theta
\left(\omega^{2}-\left(2\Delta\right)^{2}\right)}{\omega\sqrt{\omega^{2}-\left(2\Delta\right)^{2}}}.
\end{equation}
This result is, up to the $d-$wave weighted electron-pocket density
of states $\rho=\frac{1}{N}\sum_{\mathbf{k}}\gamma_{B_{1g}}^{2}\left(\mathbf{k}\right)\delta\left(\epsilon_{F}-
\epsilon_{\mathbf{k},el}\right)$,
the usual one for a fully gapped superconductor \cite{Klein}. The
square root divergence for $\omega\rightarrow2\Delta$ from above
is the well established pair-breaking peak in the electronic Raman
spectrum of a superconductor \cite{Devereaux-RMP}. For our considerations we also need the
real part of $\chi_{{\rm nem}}^{0}\left(\omega\right)$ shown in Fig. \ref{fig1}(a). The important aspect of the real
part occurs for $\left|\omega\right|<2\Delta$. In this regime it follows after Kramers-Kronig
transformation
 \begin{equation}
{\rm Re}\chi_{{\rm nem}}^{0}\left(\omega\right)=\chi_{{\rm nem,\infty}}^{0}+\rho \frac{\left(2\Delta\right)^{2}\arcsin
\left(\frac{\omega}{2\Delta}\right)}{\omega\sqrt{\left(2\Delta\right)^{2}-\omega^{2}}}.
\end{equation}
Besides the low-energy contribution, there is a weakly frequency dependent contribution from the
high energy fermions, which we approximate by a constant $\chi_{{\rm nem,\infty}}^{0}$. It can be absorbed
in the dimensionless coupling $\lambda=\rho g/\left(1-g\chi_{{\rm nem,\infty}}^{0}\right)$, and henceforth we
take $\lambda$ as the tuning parameter that increases with increasing $g$.

\par
Because of its singular behavior, the leading contribution to ${\rm Re}\chi_{{\rm nem}}^{0}\left(\omega\right)$
for frequencies below $2\Delta$ is the second term above. If we insert
this result into Eq.\ref{eq:Raman}, we find a sharp pole in the renormalized
Raman response once the condition $1=g{\rm Re}\chi_{{\rm nem}}^{0}\left(\Omega_{r}\right)$
is fulfilled for $\Omega_{r}<2\Delta$, where the imaginary part vanishes.
As illustrated in Fig.\ref{fig1}(a), the square root divergence of ${\rm Re}\chi_{{\rm nem}}^{0}\left(\omega\right)$
as $\omega$ approaches $2\Delta$ from below, guarantees that the
above condition can be satisfied and that a new exciton pole emerges.
The resonance frequency of the nematic response is
\begin{equation}
\Omega_{r}=2\Delta \sin \theta(\lambda),
\label{eq:omegares}
\end{equation}
where $\theta(\lambda)$ obeys the equation $\sin2\theta=2\lambda \theta$.

In the clean limit and at $T=0~K$, the mode is arbitrarily sharp as
it occurs below the particle-hole continuum that starts at $2\Delta$.
Near the resonance the electronic Raman response is:
\begin{equation}
R_{B_{1g}}\left(\omega\right)=Z_{r}\delta\left(\omega-\Omega_{r}\right)
\label{eq:resonance-R}
\end{equation}
with spectral weight $Z_{r}=\zeta\left(\frac{\Omega_{r}}{2\Delta}\right)$
determined by
$\zeta\left(x\right)=\frac{\pi x\left(1-x^{2}\right)\arcsin\left(x\right)}
{x\sqrt{1-x^{2}}+\left(2x^{2}-1\right)\arcsin\left(x\right)}$. $Z_{r}$ must be compared with the total weight
$\pi^2\rho\Delta/2$ of the usual pair-breaking peak of the BCS theory. Near $2\Delta$ the spectral weight
vanishes linearly $Z_{r}\approx \pi \frac{\Omega_{r}-2\Delta}{\Delta}$.
Since $\rho\Delta\ll1$, the resonance mode soon acquires a weight
comparable to the non-interacting BCS pair-breaking peak. A typical value for $\Omega_{r}=\frac{3}{4}2\Delta$
is $Z_{r}\approx 1.45$.

\par
In Fig. \ref{fig1}(b) we show the Raman response for different coupling constants.
We included a small but finite intrinsic width $\Gamma$ that
may result from impurity scattering or thermal excitations of quasiparticles \cite{SI}.
We see that for small nematic coupling constants the pair-breaking
peak keeps its line shape but increases in weight. Once $2\Delta-\Omega_{r}$
is larger than $\Gamma$, a peak that is well separated from the continuum above 2$\Delta$ emerges and sharpens with a
Lorentzian line shape for larger coupling strength. The sharpness of the Raman resonance relies heavily
on the opening of a full gap in the superconducting spectrum. Nodes of the gap around
the electron pockets essentially wash out the nematic resonance \cite{SI}, making the presence of the
nematic resonance a clear indication of the absence of nodes around the electron pockets. We also note that,
already in the
absence of any nematic correlations, disorder induces a much weaker temperature dependence of the pair-breaking
peak energy compared to the simple BCS result for $\Delta(T)$ \cite{Devereaux93},
and we expect a similar effect for the $T$-dependence of the nematic resonance energy $\Omega_{r}$.

\par
\emph{Lattice cutoff.}---
Next we discuss the fate of the nematic resonance as the system approaches the nematic QCP
by increasing $\lambda$.
From Eq.~(\ref{eq:action}) it is tempting to deduce that the QCP, defined by $g \chi^0_{\rm nem}(0) = 1$, is
given by $\lambda=1$.
This, in turn, would imply that, using Eq.~\ref{eq:omegares}, the resonance frequency  $\Omega_r \rightarrow 0$,
while the spectral weight of the resonance $Z_{r}\approx\frac{3\pi\Delta}{2\Omega_{r}}$ diverges.
\begin{figure}
\centering
\includegraphics[clip,width=0.99\linewidth]{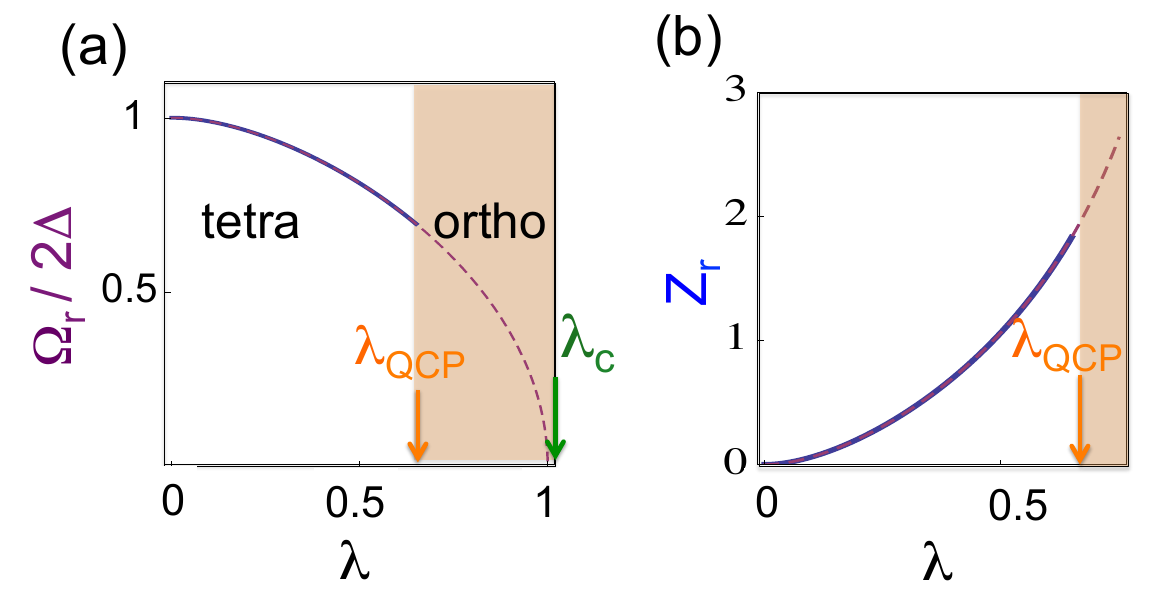}
\caption{Evolutions of the nematic resonance energy $\Omega_r$ (a) and spectral weight $Z_r$
(b) with the dimensionless coupling constant $\lambda$ (dynamic limit). Because of finite nemato-elastic coupling,
$Z_r$ does not diverge, and $\Omega_r$ does not soften completely, at $\lambda_{QCP}$ (see text).}
\label{fig2}
\end{figure}
However, in practice, this is not the case since, in addition to the purely electronic contribution of
 Eq.~(\ref{eq:action}), there is a symmetry-allowed coupling to the lattice degrees of freedom
given by \cite{footnote}
\begin{equation}
H_{c}=\gamma\int d^{d}r\phi\left(r\right)\left(\partial_{x}u^{x}-\partial_{y}u^{y}\right).
\label{eq:nemelast}
\end{equation}
$\gamma$ is a nemato-elastic coupling constant and $\mathbf{u}$
is the usual phonon displacement field. In the harmonic approximation
of the lattice, the effect of the coupling to the phonons is a renormalization
of the nematic coupling constant
\begin{equation}
g\left(\mathbf{q},\omega\right)=g+\gamma^{2}\frac{\mathbf{q}^{2}}{C_{s}^{0}\mathbf{q}^{2}-\omega^{2}}.
\end{equation}
$C_{s}^{0}$ is the bare value of the orthorhombic elastic constant and $\textbf{q}$ is a wave vector along
the critical directions of the Brillouin zone \cite{Cano10}.
Note that, the static limit $g_{{\rm stat}}\equiv g\left(\mathbf{q},0\right)=g+\frac{\gamma^{2}}{C_{s}^{0}}$$ $
and the dynamic limit $g\left(\mathbf{q}=\mathbf{0},\omega\right)=g$
do not commute \cite{Kontani}.
Thus, the enhancement of the static nematic coupling constant due
to the coupling to elastic degrees of freedom does not enter the
dynamic Raman response which filters out the purely electronic contribution of the nematic response.

On the other hand, it is the static limit that governs
thermodynamics and the actual nematic phase transition. Thus, the condition for the nematic QCP
is $\lambda_{{\rm stat}}=1$, where $\lambda_{{\rm stat}}=\rho g_{{\rm stat}}/\left(1-g_{{\rm stat}}
\chi_{{\rm nem,\infty}}^{0}\right)$ is the dimensionless nematic coupling constant in the static regime,
renormalized by high energy excitations. Since $g_{{\rm stat}}>g$, it follows $\lambda < \lambda_{{\rm stat}} \leq 1$.
In other words, even at the nematic QCP $\Omega_r$ does not soften to zero frequency, and concomitantly,
$Z_r$ stays finite (see Fig. \ref{fig2}) \cite{Note}. In the symmetry-broken nematic phase ($\lambda_{{\rm stat}} > 1$)
the behavior of $(\Omega_r, Z_r)$ is non-universal, and is beyond the scope of this work.

\par
\begin{figure*}
\centering
\includegraphics[clip,width=0.99\linewidth]{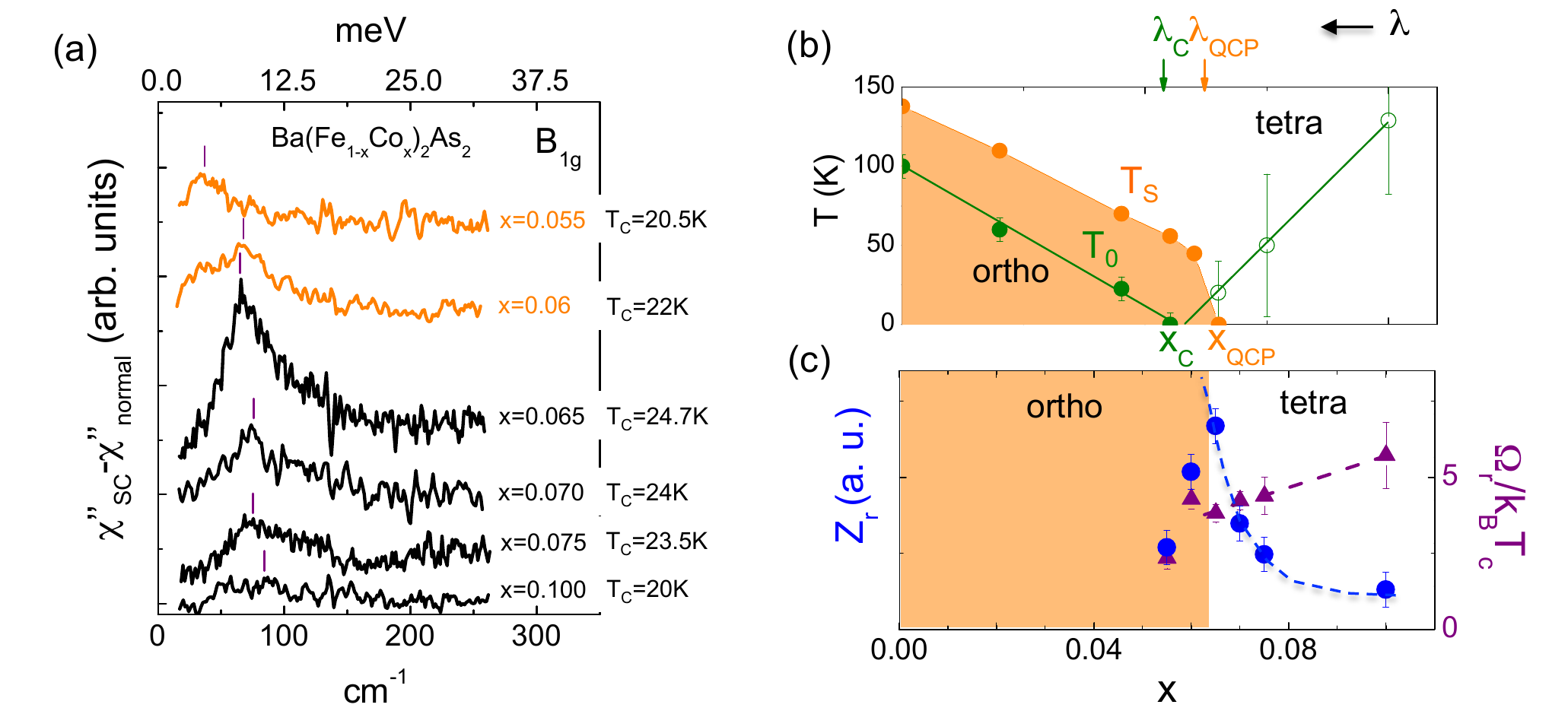}
\caption{(a) Co doping ($x$) dependance of the $B_{1g}$ Raman response in the superconducting state (SC) of
Co-Ba122 \cite{Chauviere10}. The spectra have been subtracted by those just above $T_c$ to highlight the SC
induced features. $x$=0.065 corresponds to optimal doping ($T_c$=24.7~K) and the spectra in orange correspond to $x$ compositions in the orthorhombic phase. (b) Phase diagram (T, x) of Co-Ba122 showing the evolution of $T_S$ (orange) and of the electronic nematic Curie-Weiss temperature $\left| T_0\right|$ (green) deduced from Raman experiments  in the normal state \cite{Gallais13}. Decreasing $x$ content is equivalent as increasing the coupling $\lambda$ of our theory. Note that $T_0$ is positive for $x<x_c$ (filled symbols) and negative for $x>x_c$ (open symbols). (c) Evolution of $Z_r$ and $\Omega_r$ as a function of Co doping $x$ in Co-Ba122. }
\label{fig3}
\end{figure*}

\emph{Raman Experiments.}---
We now make contact with the experiments on Fe SC and in particular with electron doped
BaFe$_2$As$_2$ (Co-Ba122) \cite{Muschler,Chauviere10,Gallais13}.
We assume that decreasing Co doping $x$ is equivalent to increasing $\lambda$.
In the superconducting state of Co-Ba112 the most
salient feature of the Raman spectra is a peak observed in $B_{1g}$ symmetry only \cite{Muschler}.
Until now the $B_{1g}$ peak has been attributed to a 2$\Delta$ BCS pair breaking peak coming from the electron
pockets \cite{Muschler,Mazin10,Chauviere10}. However as first reported in Ref. \cite{Chauviere10}, its Co
doping dependance displays a striking behavior which is shown in Fig. \ref{fig3}(a). Coming from the overdoped
tetragonal side ($x$=0.1, $T_c$=20~K) of the phase diagram, the spectral weight of the peak increases
dramatically upon approaching the orthorhombic/nematic phase which occurs below $x$=0.065. It reaches a maximum
at $x$=0.065 before collapsing in the orthorhombic phase at lower $x$. As stated above, the spectral weight of a
simple BCS Raman pair-breaking peak is expected to scale as the gap energy $\Delta$. However the strong enhancement
of $Z_r$ around optimal doping, where $T_c$ is maximum, cannot be attributed to substantial changes in $\Delta$, since this would imply equivalent changes in $T_c$, whereas, in practice the $T_c$ changes by less than 20$\%$ within the doping range considered. Moreover the peak
energy in units of k$_B$T$_c$ actually softens upon approaching $x$=0.065, going from $\sim$ 5 to $\sim$ 4~$k_BT_c$ (see Fig. \ref{fig3}(c)), in a manner consistent with the nematic resonance scenario.
\par
The above observations can be naturally explained by the nematic resonance scenario whose spectral weight
$Z_r$ is strongly enhanced by the proximity of an nematic instability. In Co-Ba122
the position of the purely electronic nematic instability $x_c$ can be determined by looking at
the enhancement of Raman $B_{1g}$ nematic fluctuations observed in the normal state of the tetragonal
phase \cite{Gallais13}. The enhancement of the associated nematic susceptibility was found to follow a Curie-Weiss temperature dependence over a wide range of Co doping from $x$=0 to $x$=0.1. As shown in Fig. \ref{fig3}(b), the extracted Curie-Weiss temperature $T_0$ is lower than $T_S$, the tetragonal to orthorhombic transition temperature, and extrapolates to T=0~K at $x_c\sim$~0.055.
In a purely electronic model the spectral weight $Z_r$ would diverge at $x_c$. However, as already stressed
above, due to finite nemato-elastic coupling, the actual quantum critical point is moved to a higher
doping $x_{QCP}$, between $x$=0.06 ($T_S$=46~K) and $x$=0.065 ($T_S$=0~K), where $Z_r$ and $\Omega_r$ are still finite as observed experimentally (Fig. \ref{fig3}(b)) .
\par
We note that a similar divergence of the $B_{1g}$ peak spectral weight has been observed recently in Co doped NaFeAs close to the
boundary between the tetragonal and orthorhombic phase \cite{Blumberg}. There the $B_{1g}$ resonance was found to be almost
Lorentzian below $T_c$, with a linewidth significantly smaller than in the case of Co-Ba122: 1~meV and 5~meV respectively near
optimal doping. This difference possibly reflects a smaller coupling constant $\lambda$ in Co-Ba122 which would make the
nematic resonance closer to the 2$\Delta$ continuum and therefore broader. Recent ARPES data in Co-Ba122 indeed
indicate that the $B_{1g}$ peak energy is only slightly below 2$\Delta$ of one of the electron pocket \cite{Hajiri14}. Disorder may also play a role as optimally doped Co-Ba122 has a significantly higher Co content than Co-Na111 (see supplemental materials for theoretical fits of the experimental lineshapes \cite{SI2}). However, we stress that, while the lineshape of the resonance can be material dependent, the key feature of our theory, namely the enhancement of the spectral weight near the nematic QCP, is observed experimentally in both systems.

\par
The nematic resonance scenario captures remarkably well the salient features of the $B_{1g}$
peak observed in the SC state of electron doped Fe SC: its nematic symmetry ($B_{1g}$) and the strong
enhancement of its spectral weight near the nematic QCP. It also makes a direct
link between the nematic response in the normal and superconducting state via Eq. \ref{eq:Raman}: close
to $x_c$ the enhanced nematic susceptibility of the normal state converts into a sharp nematic resonance
at finite energy in the SC state because of the gap opening.

Furthermore, our theory leads to the following three predictions which can be verified in future experiments.
(i) From Eq.~\ref{eq:omegares}, and assuming $\lambda$ to be $T$-independent, we expect that
$\Omega_r \propto \Delta$ upon varying temperature. (ii) Due to the nemato-elastic coupling of Eq.~\ref{eq:nemelast} the orthorhombic elastic constant is renormalized to $C_s = C_s^0 - \gamma^2 \chi_{\rm nem}(\omega =0)$. From Eq.~\ref{eq:resonance-R} and from Kramers-Kronig relation we deduce that the elastic softening $\delta C_s \equiv C_s - C_s^0$ scales as $\delta C_s \propto Z_r/\Omega_r$ upon approaching the nematic transition deep in the SC state, as a function of either doping or temperature. (iii) The nematic resonance will lead to single particle renormalizations (like peak-dip-hump features) everywhere on the electron pockets which should be visible in ARPES experiments. 

\par
\emph{Conclusion.}---
In summary we have demonstrated the generic presence of a nematic resonance in the SC state of systems near a nematic
quantum critical point like Fe SC. The very existence of this resonance relies only on the presence of a fully gapped Fermi
pocket and the proximity of a nematic quantum critical point. Existing Raman data in the $B_{1g}$ symmetry channel of several
Fe SC systems like Co-Ba122 and Co-Na111 indicate that this resonance actually dominates the Raman response in the SC state.
It provides the most striking manifestation of nematicity in the superconducting properties of Fe SC.

\acknowledgements{We are grateful to U. Karahasanovic and R. Hackl for discussions. Y.G. and L.C. acknowledge support
from the ANR grant PNICTIDES. J.S. acknowledges support from Deutsche Forschungsgemeinschaft through DFG-SPP 1458
\emph{Hochtemperatursupraleitung in Eisenpniktiden} and
thanks the Universit\'e Paris Diderot, where this work was done, for
its hospitality. }


\begin{thebibliography}{10}
\bibitem{footnote} We use a unit cell with one Fe atom. If one uses two Fe atoms per unit cell the $B_{1g}$
symmetry becomes the $B_{2g}$ symmetry and $\phi$ couple to $\partial_{x}u^{y}+\partial_{y}u^{x}$.
\bibitem{Note} If one allows for a small but finite photon momentum, the elastic contribution is still suppressed, albeit not completely, and $\lambda<\lambda_{stat}$ still holds.
\end{thebibliography}
\end{document}